\documentclass[aoas,nameyear,dvips]{arximspdf}

\doi{10.1214/10-AOAS385}
\volume{4}
\issue{4}
\pubyear{2010}
\firstpage{1644}
\lastpage{1648}

\begin{document}
\begin{frontmatter}

\title{Remembering Leo Breiman}
\runtitle{Remembering Leo Breiman}

\begin{aug}
\author{\fnms{Richard A.} \snm{Olshen}\corref{}\ead[label=e1]{olshen@stanford.edu}}
\runauthor{R. A. Olshen}
\affiliation{Stanford University}
\address{Division of Biostatistics\\
Stanford University School\\ \quad  of Medicine\\
HRP Redwood Building\\
Stanford, California 94305-5405\\
USA\\
\printead{e1}} 
\end{aug}

\received{\smonth{7} \syear{2010}}



\end{frontmatter}

I published an interview of Leo Breiman in \textit{Statistical Science}
[Olshen (\citeyear{Olshen2001})], and also the solution to a problem concerning almost
sure convergence of binary tree-structured estimators in regression
[Olshen (\citeyear{Olshen2007})]. The former summarized much of my thinking about Leo up
to five years before his death. I discussed the latter with Leo and
dedicated that paper to his memory. Therefore, this note is on other
topics. In preparing it I am reminded how much I miss this man of so
many talents and interests. I miss him not because I always agreed with
him, but instead because his comments about statistics in particular and
life in general always elicited my substantial reflection.

Technical comments here are in part my responses to Leo's 2001 paper in
\textit{Statistical Science} [Breiman (\citeyear{Breiman2001})]. The paper is interesting
and provocative, but it demonstrates an attitude that seemed somewhat
unfortunate in 2001 when it was published and remained so in 2005 when
Leo died. It is even less fortunate today. D.~R. Cox may have stated the
obvious when he noted in his discussion [Breiman (\citeyear{Breiman2001}), page~216] that,
``Like all good caricatures, it contains enough truth and exposes enough
weaknesses to be thought-provoking.''

In his discussion of the paper, Bradley Efron states (page~219) that,
``Prediction is certainly an interesting subject. Leo's paper overstates
both its role and our profession's lack of interest in it\ldots the
whole point of science is to open up black boxes, understand their
insides, and build better boxes for the purposes of mankind\ldots we can
hope that the present paper was written more as an advocacy device than
as the confessions of a born-again black boxist.''

For years I have preferred Cox's approach [Breiman (\citeyear{Breiman2001}), page~216].
``Professor Breiman takes data as his starting point. I would prefer to
start with an issue, a question or a scientific hypothesis.'' Also, I
believe strongly that crisp mathematical formulations of statistical
problems can clarify rather than obscure them; likely, if pressed Leo
would have agreed. The paper and Bruce Hoadley's discussion of it focus
on the importance of predictors. A predictor might be ``important'' if
it predicts whatever outcome is in question accurately by itself.
Alternatively, it might be called ``important'' if the performance of
other predictors is harmed by its absence. More generally, a variable
might be deemed important if it is approximately mutually predictable
with a set of predictors, and the entire set is important or not by
either criterion. These notions permit easy expression in mathematical
terms, though space precludes precise statements here. Both the paper
and much discussion of it are about selecting predictive features, in
particular about tracking the behavior of features as time, age, or some
other dimension varies. This may amount to choosing a parsimonious set
of basis functions for a linear space of functions that describes the
sample paths of that feature. Coefficients of the feature in the
carefully selected basis then become features themselves in whatever
classification or prediction is required. See, for example, Sutherland et
al. (\citeyear{Sutherland1988}), Chapter 10.

Suppose that an ``outcome'' $y$ might be predicted from input $x$, and
that the mechanism by which the outcome is determined involves not only
the input $x$, but also noise. The conditional distribution of $y$ given
$x, f_{\theta} (y|x)$ might depend also on unknown parameters; denote
them by theta ($\theta)$. There are, then, three obvious sources of
randomness. Leo argues---I think correctly---that the principal issue
facing the scientist who might make inferences from data is to predict
some future $y$* drawn from the distribution described by $f$, but not
necessarily to make statements about $f$ itself. Leo scoffed, probably
unfairly, at the substantial energy spent by members of the statistical
community quantifying information about $\theta$ available from data. He
was slightly unfair when he spoke derisively (page 204) of ``Bayesian
methods combined with Markov chain Monte Carlo.'' The strict,
frequentist approach to inference has as sources of randomness the
noise, and $x$ itself. Only in ``random effects'' models, though
conventional for a Bayesian, does the distinction become blurred; see
Hill (\citeyear{Hill1965}). The frequentist does not impute randomness to $\theta ;$
for that person it is a leap to impute randomness of any origin,
subjective or frequentistic, to it. I am surprised at the pejorative
lumping together of the different approaches to inference about ($f, x,
\theta)$, ignoring, as it were, upon what inference is conditioned. More
or less, frequentistic inference is conditioned on $\theta$ and Bayesian
on $x$. Though Leo's paper was published nearly 10 years ago, he speaks
(Section~11.3, page~213) of microarray data, which hardly permit
analyses without at least an implicit Bayesian/random effects model for
the distribution of $p$-values in rows of a large rectangular
array. For careful discussion of how Bayesian formulations, empirical
Bayes solutions, and predictions regarding future data combine to yield
much of interest regarding expression arrays and brain imaging, to cite
two areas of application, see Efron's papers on local false discovery
rates (\textit{locfdr}), available through
\url{http://stat.stanford.edu/~ckirby/brad/papers/}.

Section 11.1 and following of Leo's obviously provocative 2001 paper is
about survival analysis. His acknowledgments (page~215) make explicit
mention of his and my discussion of the celebrated Cox model in
particular and of biostatistics in general. In this arena I believe that
Leo's critique is right on. For starters, survival itself, rather than
the more difficult concept of hazard, especially relative hazard, is of
paramount interest. Predicting survival for the next patient matters far
more than does testing any hypothesis about past patients. Leo's various
complaints about the practices and orientation of some leading
statistical journals would be difficult to report diplomatically when it
comes to testing the parameters of a particular model for survival
versus predicting survival for the next patient. His were not merely the
ruminations of a senior professor at a famous university who may have
had a paper rejected. ``Testing'' parameters of a model is fraught with
difficulties that owe in part to sample size, even when the model is
correct and censoring is not an issue. This critique is applicable no
matter the philosophical underpinnings of scientific inference.
Difficulties include but are not limited to the celebrated ``Lindley
Paradox.'' See, for example, Lindley (\citeyear{Lindley1957})
and Lehmann (\citeyear{Lehmann1958}).

I was fortunate to be able to discuss with Leo work by Piette, Nazari
and Olshen (\citeyear{Piette1998}). It ran afoul of the editor of a major statistical
journal and, perhaps unwisely, was never revised for submission
elsewhere. What separates this paper from many others in survival
analysis is its prediction of readmission for future substance abuse
patients rather than inference about parameters of models that describe
past patients. It uses a parametric bootstrap applied for the most part
to Weibull models that are shown to fit the data. The bootstrap was
adapted to prediction in our context. To be precise, the paper is a case
study of readmission patterns following 42,648 discharges of patients
treated for substance abuse in U.S. Department of Veterans Affairs
hospitals for a year straddling 1990--1991. We learned that substance
abuse inpatients are at extremely high risk of readmission, particularly
if they are more than 65 years of age and have chronic medical problems.
Risk of readmission is highest immediately following discharge and
declines subsequently. We learned that intrinsic difficulties in
predicting readmission rather than limitations of our model, account for
its varying accuracy.

While Leo Breiman was certainly an important statistician, probabalist,
and colleague, he was also a good friend. Leo's notion of compromise was
clear enough to all who knew him, but it was not the modal approach.
Think of a couple, one of whom wishes to live in New York, while the
other wants Los Angeles. Though living in St. Louis might be a
compromise of sorts, it would please neither party. So it was with Leo:
your way or my way, but necessarily one of the two. If it's not always
your way or always mine, it might be said that the party is willing to
compromise. That Leo was willing to ``compromise'' was illustrated in
bringing our four-author book [Breiman et al. (\citeyear{Breiman1984})] to completion.

Leo and Chuck Stone had neither spoken for awhile nor did either have an
algorithm whereby the book could be completed, no matter that each
author in his own way had spent much effort furthering CART. One day Leo
and Jerry Friedman were having lunch in a restaurant on Hearst in
Berkeley. Per chance, Chuck Stone, my wife Susan Olshen, and I came to
the same restaurant. To ignore one another would have been to be
publicly rude, something to which Leo was quite allergic. On Leo's
urging, the four coauthors agreed to meet after lunch in his office.
Susan said that she wouldn't attend, instead would go to a library to
read a book. I said, ``Nothing doing. You come to the meeting.'' As the
meeting began, Leo took a chair in back of his desk, with Jerry seated
to one side. Chuck and I were to be seated facing Leo from the far side
of the desk. Deliberately, I took a chair with an elevated seat and
placed it adjacent to the desk, between Leo and Jerry on one side and
Chuck and me on the other. I asked Susan to sit in the carefully
situated chair. I guessed that the ever gallant Leo would be as
accommodating in that scenario as in any other, and I hoped that he and
Chuck could agree on whatever needed agreeing. Surely Jerry and I would
go along with anything to which Leo and Chuck agreed. \textit{Mirable
dictu}! Leo and Chuck came quickly to at least superficially amicable
agreement, in the Breiman style. All of us did what we decided to do,
and the book was born. The rest was downhill.

Even when dealing CART itself, occasionally Leo may not have been
sufficiently generous to prior contributions by others. He stated
[Breiman (\citeyear{Breiman2001}), page~207] that, ``While trees rate an A$+$ on
interpretability, they are good, but not great, predictors. Give them,
say, a B on prediction.'' Of course, ``boosting'' and other technologies
by which to enhance trees, many discussed by Leo in his paper, were well
known before 2001. The very early reference by Morgan and Sonquist
(\citeyear{Morgan1963}) to CART-like algorithms advertised them as, ``automatic
interaction detectors.'' See the discussion on pages~181 and~216 of
Breiman et al. (\citeyear{Breiman1984}). In work not reported here, others and I have
found that if splits on successive nodes of a binary decision tree are
taken to suggest two-factor interactions, while splits on individual
nodes have suggested main effects, then plugging the entire list of
candidates into ``the lasso'' [Tibshirani (\citeyear{Tibshirani1996})] and plugging output
into most any reasonable classifier can lead to accurate prediction.
Admittedly, validating the entire process is difficult.

In closing this remembrance I am reminded of a story told me by a former
Berkeley colleague, now sadly also deceased. It concerned appointments
of Leo and Chuck Stone to the UC Berkeley faculty. He said to the dean
who was considering the files. ``Look. You know me. I'm against almost
everything. But I think that these would be great appointments.''
Fortunately, as it turned out, Berkeley got this decision right and
appointed two remarkable individuals to its esteemed faculty. One of the
two, Leo, has been gone now for five years. It understates the case to
say that I miss him very much!

\printaddresses

\end{document}